# Fragmentation of Positronium (Ps) in collision with Li ion

S. Roy and C. Sinha

Theoretical Physics Department, Indian Association for the Cultivation of Science, Kolkata - 700032, India.

**ABSTRACT:**

Fragmentation of ground state ortho Positronium (Ps) in collision with Li ion ($Li^+$) is studied in the framework of post collisional Coulomb distorted eikonal approximation (CDEA) for the target elastic case. The present model takes account of the two center effect on the ejected $e$ which is crucial for a proper description of the projectile ionization involving an ionic target. Both the fully differential (TDCS) and the doubly differential (DDCS) cross sections (energy spectra) are investigated at intermediate and high incident energies. A broad distinct Electron loss peak (ELP) centered around $v_e \approx v_p$ is noted in the $e$ energy spectrum in contrast to the sharp ELP peak for a heavy projectile. Two salient features are noted in the present study: i) the shift of the $e$ DDCS peak (summed over $e^+$ angles) towards higher ejection energy with respect to half the residual energy of the system, ii) comparison of the $e$ & $e^+$ energy spectra reflect a strong $e$ - $e^+$ asymmetry with respect to the ratio $v_e/v_p = 1$. Both these features could be attributed to the post collisional two center effect on the $e$ due to its parent nucleus ($e^+$) and the screened target ion. Two different wave functions of the Li ion are chosen in order to test the sensitivity of the present results with respect to the choice of the wave function.

PACS No: 34.90 + q

**Introduction :**

Electron emission process in atom – atom or ion - atom collisions becomes particularly interesting at the same time complex when a structured projectile loses electron in collision with the target. Two independent channels can contribute to such projectile electron loss process e.g., the projectile electron can be knocked out by the screened target nucleus or by a target electron [ 1 ]. In the former process (singly inelastic) the target usually remains in its ground state i.e., target elastic while in the latter ( doubly inelastic) , the target gets excited or ionized i.e., target inelastic . Since these two channels lead to different final products, their contributions are to be added incoherently (i.e., in the cross section level) . The relative importance of the two channels depends on the incident energy as well as on the particular collision system.

Most of the earlier experiments [2-11] and consequently theories [12-19 ] on the projectile ionization concentrated on bare , partially stripped [ 2-6, 12, 13, 16-19 ] or neutral heavy projectiles [ 7-11, 14, 15 ] for which a cusp shaped peak was observed in the emitted electron energy spectrum at around $v_e$ ( velocity of the electron ) $\approx v_p$ ( velocity of the positron ). This peak was attributed to the electron loss from the projectile ion / atom into its low - lying continuum, usually referred to as the ELP peak . With the advent of mono energetic energy tunable positronium (Ps) beams [ 20, 21 ] , attention is also being focused both experimentally [ 22- 24 ] and theoretically [ 25-28 ] on the breakup process of the projectile Ps atom .

The basic difference between the heavy projectile and the light projectile fragmentation is that, in the former case the deflection as well as the energy loss of the projectile, due to its heavy mass is negligibly small leading to a pronounced peak / cusp in the forward direction, while in the latter case, the light projectile can scatter to large angles and its energy loss is also not negligible leading to a broad peak / cusp. Study of the dynamics e.g., angular and energy distributions of the process gives valuable information about the ionizing mechanisms and provides a unique insight into the collision dynamics as well as the atomic structure of the collision partners.

The Ps atom is now considered to be an ideal probe to solid surfaces for determining their structures mainly because it can only undergo elastic reflection from the outer surface layer of a solid [29]. Because of the large break up probability of the Ps (above its binding energy 6.8 eV), the multiple scattering effect from inner – layer atoms of the solid is expected to be negligible for the Ps (unlike the low energy electron and positron) and as such the low energy Ps collision should be confined to the outer most surface layers. However, neutral atoms and molecules like He, $H_2$ also interact mainly with the surface atoms, but the available low energy beams are not energetic enough to probe small scale surface structure [29,30]. Thus the Ps provides a great deal of advantage over charged as well as neutral heavy projectiles as a probe to study the structure of atoms and molecules and the surface properties of solids and the knowledge of different scattering parameters for Ps – atom, molecule or ion collisions could be highly useful for such studies. Further, by virtue of very light mass of Ps (3 orders of magnitude smaller than hydrogen), its interaction with various forms of matter, ranging from electrons, protons, alkali ions to atoms, molecules, solid surfaces and plasmas can provide important information about the target medium.

In an earlier work [26] we studied the breakup of Ps in collision with a hydrogenic ($He^+$) ion where the results showed prominent signature of the two center effect on the ejected electron in the final channel. The present work addresses the extension of it [26] to a two electron (helium like) ionic target, $Li^+$ for the target elastic case with an intention to confirm the two center effect in the case of an ionic target,

i.e., 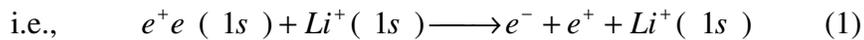

$$e^+e\ (1s) + Li^+(1s) \longrightarrow e^- + e^+ + Li^+(1s) \qquad (1)$$

One major advantage of the $Li^+$ target (over the $He^+$) is that the electrons of the former are much more tightly bound than the electron of the Ps atom and as such the probability of the electron loss from the projectile Ps is expected to be much higher than the ionization of the target. Further, we have neglected any virtual or real excitation of the $Li^+$ target during the fragmentation (i.e., target elastic case). This assumption seems to be quite legitimate in view of the large excitation energy of the target $Li^+$ as compared to the projectile Ps.

Since the initial components of the reaction (1) are both composite bodies, the theoretical prescription of such a process is rather complicated. As such one has to resort to some simplifying assumptions for the theoretical modeling of such a many body ( five body) reaction process . The present calculation is performed in the frame work of the post collisional Coulomb distorted eikonal approximation (CDEA) taking account of the proper asymptotic three body boundary condition in the final channel, which is one of the important criteria for a reliable estimate of the ionization cross sections.

**Theory**:

The prior form of the ionization amplitude for the aforesaid process (1) is given as:
$$T_{if}^{prior} = \left\langle \Psi_f^-(\vec{r}_1, \vec{r}_2, \vec{r}_3, \vec{r}_4) \right| V_i \left| \psi_i(\vec{r}_1, \vec{r}_2, \vec{r}_3, \vec{r}_4) \right\rangle \tag{2}$$

where $\vec{r}_1$, $\vec{r}_2$, $\vec{r}_3$ and $\vec{r}_4$ are the position vectors of the positron and the electron of the positronium atom and the bound electrons of the $Li^+$ ion respectively, with respect to the target nucleus.

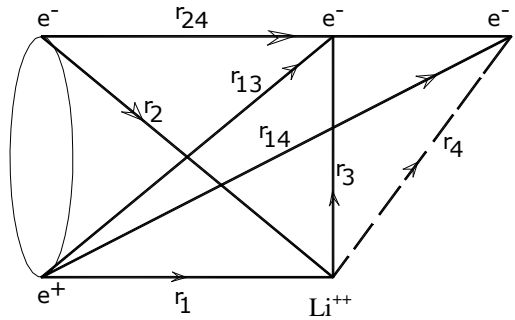

$V_i$ is the initial channel perturbation, can be given by;
$$V_i = \frac{Z_t}{r_1} - \frac{Z_t}{r_2} - \frac{1}{r_{13}} + \frac{1}{r_{23}} - \frac{1}{r_{14}} + \frac{1}{r_{24}} \tag{3}$$

$Z_t$ (= 3) being the charge of the target nucleus; $\vec{r}_{13} = \vec{r}_1 - \vec{r}_3$ , $\vec{r}_{23} = \vec{r}_2 - \vec{r}_3$, $\vec{r}_{14} = \vec{r}_1 - \vec{r}_4$ ,

$$\vec{r}_{24} = \vec{r}_2 - \vec{r}_4$$

The initial asymptotic state $\psi_i$ in equation (2) is chosen as:

$$\psi_i = \phi_{Ps}(|\vec{r}_1 + \vec{r}_2|) \, e^{i\vec{k}_i \cdot \vec{R}} \, \phi_{Li^+}(\vec{r}_3, \vec{r}_4) \tag{4a}$$

where $\vec{R} = \dfrac{\vec{r}_1 + \vec{r}_2}{2}$ and $\vec{k}_i$ is the initial momentum of the Ps atom with respect to the target nucleus. The ground state wave function of the Ps atom $\phi_{Ps}(|\vec{r}_1 - \vec{r}_2|)$ is given by;

$$\phi_{Ps}(|\vec{r}_1 - \vec{r}_2|) = N_{Ps} \exp(-\lambda_{Ps} \, r_{12}) \tag{4b}$$

with $N_{Ps} = \lambda_{Ps}^{3/2}/\sqrt{\pi}$ and $\lambda_{Ps} = 1/2$

Two different forms of the ground state wave function [ 34-36] for $Li^+$ ion are chosen :

i) due to Morse etal [ 31, 32 ] :

$$\phi_{Li^+}(r_3, r_4) = u(r_3) \, u(r_4) \tag{5a}$$

where, $u(r) = \sqrt{\lambda_{Li^+}^3 / \pi} \, \exp[-\lambda_{Li^+}(r)]$ \hfill (5b)

with $\lambda_{Li^+} = (Z_t - 1) + 0.6875$ \hfill (5c)

ii) due to Clementi Roetti ( CR ) [ 33 ] :

$$\phi_{Li^+}(r_3, r_4) = u(r_3) \, u(r_4) \tag{6a}$$

with

$$u(r) = \frac{1}{\sqrt{4\pi}} [\, N_1 C_1 \exp(-2.45055 \, r) + N_2 C_2 \exp(-4.57259 \, r) + N_3 C_3 \exp(-6.67032 \, r)] \tag{6b}$$

where, $N_1 = 7.672296$, $N_2 = 19.55701$, $N_3 = 34.454821$, \hfill (6c)
$C_1 = 0.89066$, $C_2 = 0.12328$, $C_3 = 0.00088$

To construct the final channel wave function one should note that the ejected electron is in the combined (attractive) fields of the two positive ions, e. g., its parent ion the $e^+$ and the target ion, the Li$^+$. The present model takes account of this two center effect and the final state wave function $\Psi_f^-$ in eqn. (2) satisfying the incoming wave boundary condition is approximated by the following ansatz in the framework of coulomb modified eikonal approximation:

$$\Psi_f^-(\vec{r}_1, \vec{r}_2, \vec{r}_3, \vec{r}_4) =$$

$$(2\pi)^{-3} N_{Li^+} \exp(-\lambda_{Li^+} r_3) \exp(-\lambda_{Li^+} r_4) e^{i\vec{k}_1 \cdot \vec{r}_1} e^{i\vec{k}_2 \cdot \vec{r}_2} {}_1F_1(-i\alpha_2, 1, -i(k_2 r_2 + \vec{k}_2 \cdot \vec{r}_2))$$

$$\times \exp\left\{ i\eta_1 \int_z^\infty (\frac{1}{r_1} - \frac{1}{r_{12}}) \, dz' \right\} \qquad (7)$$

with $\alpha_2 = -\dfrac{(Z_t - 2)}{k_2}$ and $\eta_1 = \dfrac{(Z_t - 2)}{k_1}$; $\vec{k}_1, \vec{k}_2$ being the final momenta of the scattered positron and the ejected electron with respect to the target nucleus respectively.

In view of Eqns. (2) - (5) we get the break up amplitude

$$T_{if} \approx -\frac{\mu_f}{2\pi} \iiint N_{Li^+} \exp(-\lambda_{Li^+} r_3) e^{-i\vec{k}_1 \cdot \vec{r}_1} e^{-i\vec{k}_2 \cdot \vec{r}_2} {}_1F_1(i\alpha_2, 1, i(k_2 r_2 + \vec{k}_2 \cdot \vec{r}_2))$$

$$\times (r_1 + z_1)^{i\eta_1} (r_{12} + z_{12})^{-i\eta_1} \left[ \frac{Z_t}{r_1} - \frac{Z_t}{r_2} - \frac{1}{r_{13}} + \frac{1}{r_{23}} - \frac{1}{r_{14}} + \frac{1}{r_{24}} \right]$$

$$\times N_{Ps} \exp(-\lambda_{Ps} r_{12}) N_{Li^+} \exp(-\lambda_{Li^+} r_4) \, d\vec{r}_1 \, d\vec{r}_2 \, d\vec{r}_3 \, d\vec{r}_4 \qquad (8)$$

After much analytical reduction [ 34, 35 ] the break up amplitude $T_{if}$ in eqn. (8) is finally reduced to a two dimensional numerical integral [ 36 ]. The triple differential cross section (TDCS) for the break up process is given by:

$$\frac{d^3\sigma}{dE_2 \, d\Omega_1 \, d\Omega_2} = \frac{k_1 k_2}{k_i} | T_{if} |^2 \qquad (9)$$

The corresponding expression for the double differential cross section (DDCS) is obtained by integrating the TDCS in Eqn. ( 9 ) over the solid angles of the $e^-$ or the $e^+$ and is given by:

$$\frac{d^2\sigma}{dE_2 d\Omega_1 (d\Omega_2)} = \frac{k_1 k_2}{k_i} \int | T_{if} |^2 \, d\Omega_2 \, (d\Omega_1) \qquad (10)$$

**Results and Discussions:**

We have computed the fully ( triple ) differential cross sections ( TDCS ) as well as the doubly differential cross sections ( DDCS ) for the ionization of Ps atom in collision with the helium like ionic target ( e.g., $Li^+$ ) for the target elastic case. Since the present study is being made in coplanar geometry, i.e., $\vec{k}_i$, $\vec{k}_1$ and $\vec{k}_2$ all being in the same plane, the azimuthal angles $\phi_1$ and $\phi_2$ can assume values, $\phi_1 = 0^0$, $\phi_2 = 0^0$ and $180^0$. For the TDCS curves, we have adopted the following conventions for the ejected angles ( $\theta_2$, $\phi_2$ ): for ( $\theta_2$, $0^0$ ) we have denoted by $-|\theta_2|$ ( recoil region ) while the angles ( $\theta_2$, $180^0$ ) are plotted as $|\theta_2|$ (binary region).

Figure 1 display the triple differential cross sections ( TDCS ) against the ejected electron angle ( $\theta_2$ ) at an incident energy $E_i = 50\,\text{eV}$, with some selected values of the scattering angle (e.g., $\theta_1 = 0^0$, $20^0$, $30^0$, $45^0$) of the positron. The TDCS ( fig.1) exhibits a broad ELP for forward emission ( $0^0$ ) of both the $e$ & the $e^+$ in contrast to the sharp ELP cusp around $0^0$ for heavy ion impact [ 11 ] . This could be attributed to the probability of deflection of the light particle ($e^+$) to higher angles in contrast to the heavy projectile which, due to its heavy mass is predominantly scattered in the forward direction( $0^0$ ). Further, as is evident from fig. 1 that the TDCS exhibits some asymmetry in the shape of the binary and recoil regions with increasing scattering angles and the cross section decreases throughout the angular region with increasing scattering angle. A prominent double peak structure appears ( in the binary region ) particularly at $\theta_1 = 45^0$, the explanation of which will be discussed below.

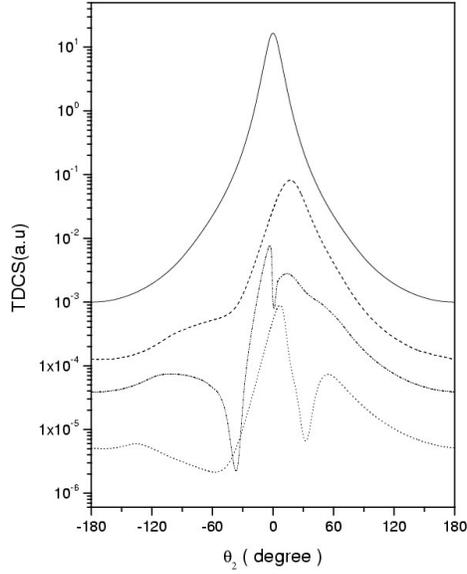

**Fig. 1.** The triple differential cross sections (TDCS) against the ejected electron angle ($\theta_2$) for different values of the scattered positron angle ($\theta_1$). The incident energy ($E_i$) is fixed at 50 eV, ejected electron energy ($E_2$) is fixed at 17 eV. The solid curve for $\theta_1 = 0^0$, dashed curve for $\theta_1 = 20^0$, dashed dot - dot curve represents $\theta_1 = 30^0$, and the dotted curve for $\theta_1 = 45^0$.

For the confirmation of the above behaviour, we have presented in fig. 2 the TDCS vs $\theta_2$ for $\theta_1 = 45^0$ at different incident energies. The occurrence of the distinct double peak particularly at $45^0$ could be attributed to the higher order effects considered in the present model. This could also be associated with the famous Thomas (p – n – e) mechanism ( TM ) [ 37, 38 ] in charge transfer problems at high incident energies. Fig. 2 also reflects that the secondary peak becomes more and more prominent with increasing incident energy, indicating the importance of the higher order effects at higher incident energies.

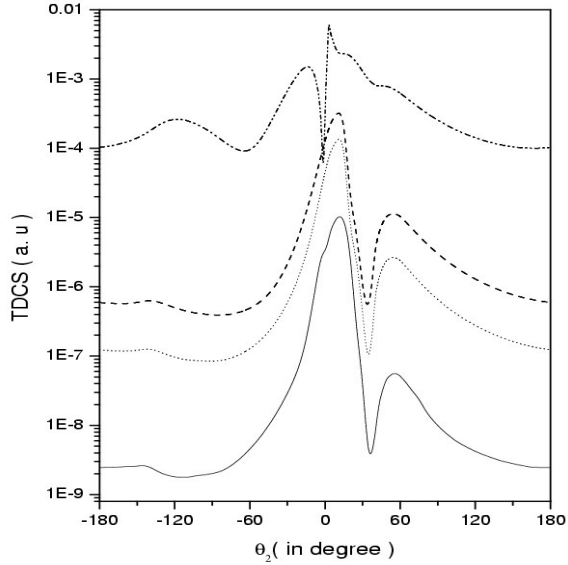

**Fig.2.** TDCS against the ejected electron angle ($\theta_2$) for different incident energies but for $\theta_1 = 45^0$. Dashed double dot curve for $E_i$ = 25eV, $E_2$ = 8 eV, dashed curve for $E_i$ = 75eV, $E_2$ = 28 eV dotted curve for $E_i$ = 100eV, $E_2$ = 40 eV and solid curve for $E_i$ = 200eV, $E_2$ = 90 eV.

Fig. 3 demonstrates the ejection energy distribution ( TDCS ) of the $e$ at an incident energy $E_i$ = 200 eV for different combinations of the ejection and scattering angles. The azimuthal angles are chosen as $\phi_1 = \phi_2 = 0^0$ i. e., when the scattered positron and ejected electron emerge on the same side of the incident beam. Following important features are noted in fig. 3 which gives a clear indication of the post collisional two center effect [11].

i ) The energy spectrum (TDCS) exhibits a broad peak (unlike the heavy projectile ) at slightly above half of the residual energy ( $E_{res}/2$ ) with some exception ( dashed dotted curve ) to be described below. The broadness being decreased with decreasing ejection angles and magnitude of the TDCS peak is enhanced with decreasing ejection angles, as is expected physically.

ii) When the $e^+$ angle ($\theta_1$) is greater than the $e$ angle ( $\theta_2$ ) the TDCS peak shifts towards the higher ejection energy with respect to $E_{res}/2$. While, the reverse behaviour is noted when $\theta_2 > \theta_1$, i. e., the peak shifts towards lower energy in this case. This shifting of the TDCS peak could be attributed to the post collisional two center effect on the ejected $e$ due to the $e^+$ and the target nucleus. Higher ejection angle ($\theta_2$) of $e$ corresponds to lower ejection energy ( $E_2$ ) while higher scattering angle ($\theta_1$ ) of $e^+$ corresponds to lower $e^+$ energy ($E_1$) i. e., higher $E_2$ , as the residual energy $E_{res}$ is shared by the $e^+$ and the $e$ if the recoil of the target ion is neglected.

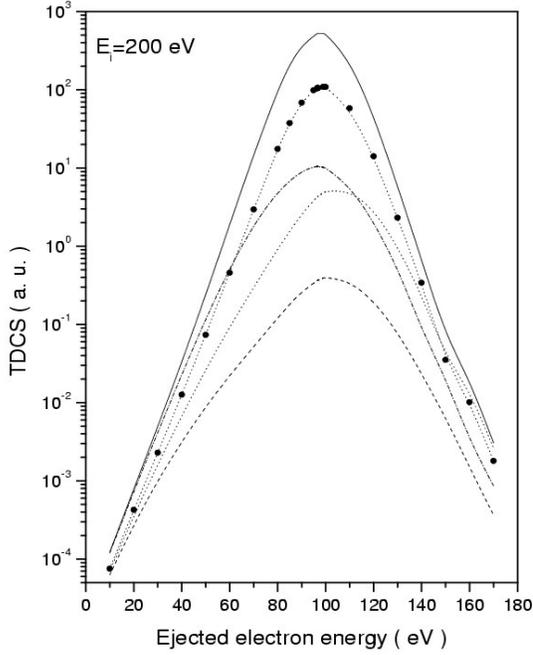

**Fig. 3.** TDCS against the ejected electron energy ( $E_2$ ) for $E_i$ = 200 eV. Solid curve for $\theta_1 = \theta_2 = 0^0$, dotted curve with solid circles for $\theta_1 = -\theta_2 = 5^0$, dash dot curve for $\theta_1 = 0^0$, $\theta_2 = 5^0$, dotted curve for $\theta_1 = 5^0$, $\theta_2 = 0^0$ and dashed curve for $\theta_1 = \theta_2 = 5^0$.

In order to test the sensitivity of the present results with respect to the choice of the target wave functions we have depicted in fig. 4 the TDCS corresponding to the two different wave functions [31 – 33] for the target Li$^+$ ion at a low ( 18 eV ) and a high ( 100 eV ) incident energies.

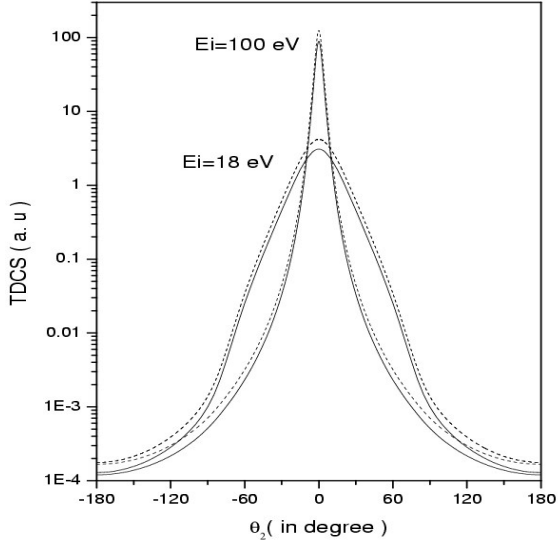

**Fig. 4.** Triple differential cross sections ( TDCS ) using different ground state wave functions of Li ion at $E_i$ = 18 & 100 eV . Solid lines for simple Morse [ 31, 32 ] wave function, dashed lines for CR wave functions [ 33 ].

The difference between the two curves is more prominent at lower incident energy as well as at backward angles ( the CR being always higher ) and tend to die out with increasing energy so that at $E_i$ = 100 eV the deviation is almost negligible, particularly at forward angles. Since the computation using the CR wave function [ 33 ] is more time consuming, we have mostly used the simpler Morse wave function [ 31, 32 ] particularly for higher incident energies where the sensitivity with respect to the wave function is quite small. Some numerical values of the fragmentation TDCS using the CR wave function [ 33] are tabulated in table I .

Table I – Differential cross sections ( in a. u. ) using the CR [ 33 ] wave function . The values within brackets indicate the power of 10.

| Angle | Energy ( eV ) | | | |
|---|---|---|---|---|
| ( degree ) | 25 | 50 | 100 | 200 |
| 0 | 1.04[+1] | 4.31[+1] | 1.73[+2] | 7.12[+2] |
| 10 | 6.19[+0] | 7.73[+0] | 3.86[+0] | 1.28[+0] |
| 20 | 2.19[+0] | 1.08[+0] | 3.43[-1] | 1.08[-1] |
| 30 | 7.75[-1] | 2.69[-1] | 7.98[-2] | 2.55[-2] |
| 40 | 3.05[-1] | 9.39[-2] | 2.83[-2] | 9.13[-3] |
| 50 | 1.29[-1] | 4.06[-2] | 1.27[-2] | 4.16[-3] |
| 60 | 5.76[-2] | 2.01[-2] | 6.62[-3] | 2.23[-3] |
| 70 | 2.54[-2] | 1.08[-2] | 3.84[-3] | 1.33[-3] |
| 80 | 1.14[-2] | 6.15[-3] | 2.41[-3] | 8.64[-4] |
| 90 | 6.58[-3] | 3.73[-3] | 1.59[-3] | 5.98[-4] |
| 100 | 3.52[-3] | 2.51[-3] | 1.13[-3] | 4.38[-4] |
| 110 | 2.39[-3] | 1.82[-3] | 8.52[-4] | 3.38[-4] |
| 120 | 1.76[-3] | 1.41[-3] | 6.75[-4] | 2.74[-4] |
| 130 | 1.37[-3] | 1.14[-3] | 5.59[-4] | 2.29[-4] |
| 140 | 1.12[-3] | 9.62[-4] | 4.81[-4] | 2.01[-4] |
| 150 | 9.67[-4] | 8.48[-4] | 4.29[-4] | 1.80[-4] |
| 160 | 8.71[-4] | 7.76[-4] | 3.96[-4] | 1.67[-4] |
| 170 | 8.19[-4] | 7.35[-4] | 3.78[-4] | 1.60[-4] |
| 180 | 8.02[-4] | 7.22[-4] | 3.72[-4] | 1.58[-4] |

Fig. 5 demonstrates the electron energy distribution ( DDCS ) at an incident energy $E_i = 50$ eV where the DDCS refers to the summation over the positron scattering angles ($\Sigma \theta_1$, $\phi_1$, hence forth referred as electron DDCS ) for different ejection angles $\theta_2$. The inset exhibits the corresponding $e^+$ energy spectrum of the same DDCS ( i. e., $\Sigma \theta_1$, $\phi_1$ ). The DDCS corresponding to the CR wave function [33] for $0^0$ emission of $e$ is also included in fig. 5. As is apparent from the figure, the DDCS exhibits a broad peak ( like the TDCS one ), slightly shifted towards higher ejection energy with respect to $E_{res}/2$. However, this shift decreases and moves towards $E_{res}/2$ with increasing ejection angle ( $\theta_2$ ). This is because the lower energy electron is preferentially ejected at higher angles ( $\theta_2$ ) so that finally at $\theta_2 = 30^0$, the peak occurs exactly at $E_{res}/2$ ( i. e., 21.6 eV ). The broadness of the DDCS peak clearly reveals that the positron in the final state could also suffer some deflections to larger angles, apart from the dominant forward scattering ($\theta_1 = 0^0$). This is in contrast to the heavy projectile which due to its heavy mass is scattered only through forward angles giving rise to a sharp cusp in the ejected electron energy spectrum. The shifting of the DDCS peak ($\Sigma \theta_1$, $\phi_1$) towards higher $e$ ejection energy could be attributed to the same fact ( two center effect ) described for the TDCS.

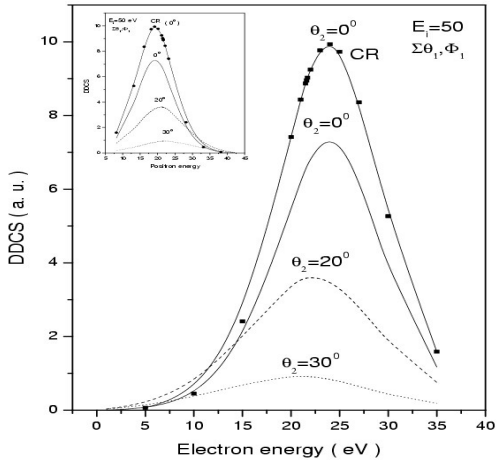

**Fig. 5**. The electron DDCS ( $\Sigma\theta_1, \phi_1$ ) against the ejected $e$ energy for different values of the ejected $e$ angle ($\theta_2$) at the fixed incident energy ($E_i$ = 50 eV) . The solid curve is for $\theta_2 = 0^0$, dashed curve is for $\theta_2 = 20^0$ and the dotted curve is for $\theta_2 = 30^0$. Solid curve with circle corresponds to CR wave function [33] for $\theta_2 = 0^0$. Both the curves for $\theta_2 = 0^0$ are scaled down by 5 factor. The inset describes the same DDCS as in fig.5 but against the ejected $e^+$ energy.

As regards the sensitivity of the DDCS with respect to the target wave function, the qualitative behaviour of the two curves ( CR and Morse wave functions ) are more or less similar in nature ( vide fig. 5) , although a significant quantitative difference is noted particularly between the two peak values.

The positron DDCS summed over the electron ejection angles ($\Sigma \theta_2, \phi_2$) is displayed in fig. 6 against the $e$ energy , while the inset represents the same against the $e^+$ energy . In this case, the DDCS peak shifts in the reverse direction ( c.f . fig.5 & inset) , e. g., at slightly below $E_{res}/2$ for forward $e^+$ scattering ( $\theta_1 = 0^0$). Further, in contrast to fig. 5 , the peak shifts towards higher ejection energy for larger positron angle ($\theta_1$). This behaviour could be explained physically as follows . The $e^+$ DDCS ( fig 6 ) includes the contributions from all the higher ejection angles ($\theta_2$) of the $e$, apart from the dominant forward emission ($\theta_2 = 0^0$ ) and higher emission angle corresponds to lower emission energy. Further, the lower $e^+$ energy corresponds to higher $e$ energy and vice versa.

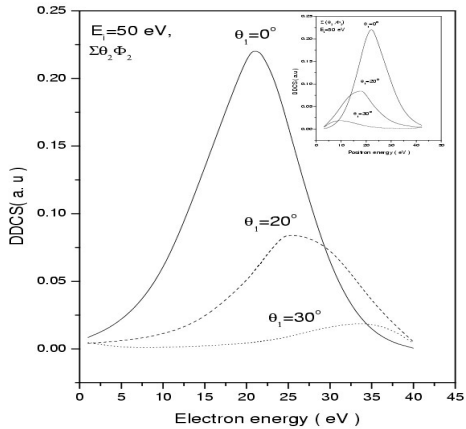

**Fig. 6.** The positron DDCS ($\Sigma\theta_2, \phi_2$ ) against the ejected $e$ energy for different values of the scattered $e^+$ angle ($\theta_1$) at the fixed incident energy ($E_i$ = 50 eV). The solid curve is for $\theta_1 = 0^0$, dashed curve is for $\theta_1 = 20^0$ and the dotted curve is for $\theta_1 = 30^0$. Solid curve is scaled down by 10 factor. The inset describes the same DDCS as in fig.5 but against the ejected $e^+$ energy.

Finally fig.7 displays the positron DDCS against the ratio $R = v_e/v_p$ for forward scattering $(\theta_1 = 0^0)$ at three different incident energies. Fig. 7 clearly demonstrates that the DDCS peak occurs at a lower ratio of $v_e/v_p$ than unity ( i. e., $v_e/v_p < 1$ ). Further, the position of the peak shifts gradually towards a higher value of R with increasing incident energy, e. g. , R being 0.97 at $E_i$ = 50 eV and attains a value of almost unity ( R ~ 1.006 ) at $E_i$ = 100 eV ( vide fig. 7 ) . A plausible physical explanation for the above behaviour could be as follows. In the post collisional interaction, the $e$ and the $e^+$ are distorted by their increasing interactions with the target. Since the $e^+$ feels repulsion while the $e$ feels attraction due to the short range interaction with the target nucleus, on an average the $e$ remains closer to the target while the $e^+$ moves away from it. As such, the probability of the $e$ ( $e^+$) to suffer hard ( soft ) collisions with the target increases with decreasing incident energy. Thus in the post collisional effect , the electron is in the combined field of its parent ( $e^+$ ) and the target nucleus.

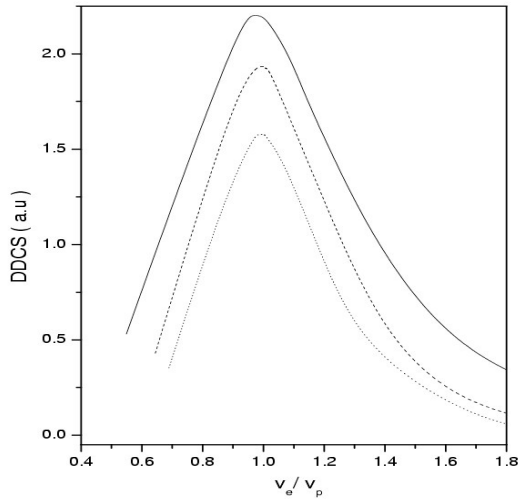

**Fig. 7.** The positron DDCS $(\Sigma\theta_2, \phi_2)$ is plotted against the ratio of $v_e/v_p$ for three different incident energies. The scattered positron angle $\theta_1$ is fixed at $0^0$. The solid curve is for $E_i$ = 50 eV, dashed curve is for $E_i$ = 75 eV and the dotted curve is for $E_i$ = 100 eV.

**Conclusions**:

The salient features of the present study are outlined below:

1. The angular distribution of the $e^-/e^+$ (TDCS) exhibits a sharp ELP at around half the residual energy ($E_{res}/2$) for forward emissions of the $e^+$ and the $e^-$ ($\theta_1 = \theta_2 = 0^0$).

2. The peak exhibits a spread in the velocity space around $\vec{v}_e = \vec{v}_p$ for forward emission of the $e^-/e^+$, unlike the sharp cusp around $0^0$ for heavy ion impact. The reason being, in contrast to the case of heavy ion impact, the $e^-$ and the $e^+$ due to their light mass, could suffer considerable deflection from their initial velocities.

3. The occurrence of a distinct double peak in the $e^-$ ($\theta_2$) distribution (exactly at $\theta_1 = 45^0$) as well as its becoming more and more prominent with increasing incident energy could be attributed to higher order effects and could possibly be associated with the famous Thomas mechanism.

4. A notable shift of the DDCS peak in the $e^-$ DDCS ( to higher momenta ) from its standard position ( $\vec{v}_e = \vec{v}_p$) is obtained. The $e^+$ DDCS shows exactly the reverse behaviour, as expected.

5. The position of the $e^+$ DDCS peak shifts gradually towards higher value of the ratio $v_e/v_p$ with increasing incident energy.

6. The sensitivity of the ground state wave function of $Li$ ion is more significant at low incident energies.

**References:**

1. D. R. Bates and G .W. Griffing  Proc. Phys. Soc. London Sect. A **67,** 663 (1954).

2. G. B. Crooks and M. E. Rudd  Phys. Rev. Lett. **25,** 1599 (1970).

3. D. Burch, H. Wieman and W. B Ingalls  Phys. Rev. Lett. **30,** 823 (1973).

4. W. E Wilson and L. H Toburen Phys. Rev. A **. 7,** 1535 (1973).

5. M. Breinig et al  Phys. Rev. A. **25,** 3015 (1982).

6. L.Gulyas, Gy Szabó, A. Kövér, D. Berényi, O. Heil and K. O. Groeneveld Phys. Rev. A. **39,** 4414 (1989).

7. H. Trabold, G. M. Sigaud, D.H Jakubassa –Amundsen, M. Kuzel, O. Heil and K. O. Groeneveld  Phys. Rev. A. **46,** 1270 (1992).

8. O .Heil, R. D. Dubois, R .Maier, M .Kuzel and K. O. Groeneveld  Phys. Rev. A. **45,** 2850 (1992).

9. M. Kuzel, R. Maier, O. Heil, D. H. Jakubassa –Amundsen, M. W. Lucas and K O Groeneveld Phys. Rev. Lett. **71,** 2879 (1993).

10. M. Kuzel, L. Sarkadi, J. Pálinkás, P. A Závodszky, R. Maier, D. Berényi, K. O. Groeneveld  Phys. Rev. A. **48** R1745 (1993).

11. M .B Shah, C. McGrath, Clara Illescas, B. Pons, A. Riera, H. Luna, D. S. F Crothers ', S. F. C. O Rourke and H. B. Gilbody Phys. Rev. A. **67 ,** 010704 ( R) (2003).

12. J. S. Briggs and F. Drepper  J. Phys. B: **11,** 4033 (1978).

13. J. S. Briggs and M. H. Day  J. Phys. B: **13,** 4797 (1980).

14. H. M. Hartley and H. R. J Walters  J. Phys. B: **20,** 1983 (1987).

15. H. M. Hartley and H. R. J. Walters  J. Phys. B: **20,** 3811 (1987).

16. H. Atan, W. Steckelmacher and M .W. Lucas  J. Phys B : **23,** 2579 (1990).

17. D .H Jakubassa –Amundsen  J. Phys B : **23,** 3335 (1990).

18. J. Wang, O. Reinhold Carlos and Joachim Burgdörfer  Phys. Rev A**. 44** 7243 (1991).

19. I. Gulyás, L. Sarkadi, J. Pálinkás, A. Kövér, T. Vajnai, Gy. Szabo, J. Vegh, D. Berenyi and S. B. Elston Phys. Rev. A. **45,** 4535 (1992).

20. A. J. Garner, A. Özen and G. Laricchia *J. Phys B*. **33** 1149 (2000).

21. A. Özen, A.J. Garner, and G. Laricchia, Nucl. Instrum. Methods Phys. Res. B. **171,** 172 (2000).

22. S. Armitage, D. E. Leslie, A. J Garner and G. Laricchia Phys. Rev. Lett **89,** 173402 (2002).

23. G. Laricchia, S. Armitage, D E Leslie Nuclear Instruments and Methods in Physics Research B. **221** 60 (2004).

24. S. Armitage, J. Beale, D. Leslie, G. Laricchia Nuclear Instruments and Methods in Physics Research B. **233** 88 (2005).

25. L. Sarkadi Phys Rev A. **68**, 032706 (2003).

26. S. Roy, D. Ghosh and C. Sinha J. Phys. B: **38** 2145 (2005).

27. C Starrett, Mary T McAlinden and H R J Walters Phys. Rev. A. **72**, 012508 (2005).

28. H R J Walters, C Starrett, Mary T Mc.Alinden Nuclear Instruments and Methods in Physics Research B. **247,** 111(2006).

29. K. F Canter in Positron Scattering in Gases, edited by Humberston J W and. Mc. Dowell M R C ( Plenum, New York, 1984 ) p - 219.

30. M. H. Weber, S .Tang, S. Berko, B. L. Brown, K. F. Canter, K. G. Lynn, A P Mills, Jr., L. O. Roellig and A .Viescas, Phys. Rev. Lett **61,** 2542 (1988).

31. P M Morse, L A Young and E S Haurwitz Phys. Rev **48**, 948 (1935).

32. B. L Moiseiwitsch and S. J. Smith Rev Mod. Phys **40,** 238 (1968).

33. E .Clementi and C. Roetti, Atomic Data and Nuclear Data Tables ( Academic, New York **14**, 268 (1974).

34. R. Biswas and C. Sinha C Phys. Rev. A. **50,** 354 (1994).

35. B. Nath and C. Sinha Eur. Phys. J. D. **6,** 295 (1999).

36. B. Nath and C. Sinha J.Phys B:**33,** 5525 (2000).

37. L. H. Thomas Proc.R. Soc. London, Ser. A **114,** 561 (1927).

38. D. Ghosh and C. Sinha   Phys. Rev. A **68,** 062701 (2003) .